# *Google Spain v. Gonzáles*:

# Did the Court forget about freedom of expression?

**Stefan Kulk and Frederik Zuiderveen Borgesius**

Contact: frederikzb@cs.ru.nl

## I. Introduction

When reviewing a job application letter, going on a first date, or considering doing business with someone, the first thing many people do is entering the person's name in a search engine. A search engine can point searchers to information that would otherwise have remained obscure.

If somebody searched for the name of Spanish lawyer Mario Costeja González, Google showed search results that included a link to a 1998 newspaper announcement implying he had financial troubles at the time. González wanted Google to stop showing those links and started a procedure in Spain. After some legal wrangling, the Spanish Audiencia Nacional (National High Court) asked the Court of Justice of the European Union (CJEU) for advice on the application of the Data Protection Directive,[1] which led to the controversial judgment in *Google Spain*.[2] In its judgment, the CJEU holds that people, under certain conditions, have the right to have search results for their name delisted. This right can also extend to lawfully published information.

---

[1] European Parliament and Council Directive 95/46/EC on the protection of individuals with regard to the processing of personal data and on the free movement of such data, OJ 1995 L 281/31.
[2] Case C-131/12, Google Spain SL and Google Inc. v. Agencia Española de Protección de Datos and Mario Costeja González, not yet published.

In this note we mainly focus on the judgment's implications for freedom of expression. Other aspects, such as the jurisdictional aspects, are only briefly touched upon in this note.[3] First, in sections II and III, the facts of the case and the CJEU's judgment are summarised. In section IV, we argue that the CJEU did not give enough attention to the right to freedom of expression. By seeing a search engine operator as a controller regarding the processing of personal data on third party web pages, the CJEU assigns the operator the delicate task of balancing the fundamental rights at stake. However, such an operator may not be the most appropriate party to balance the rights of all involved parties, in particular in cases where such a balance is hard to strike. Furthermore, it is a departure from human rights doctrine that according to the CJEU privacy and data protection rights override, "as a rule", the public's right to receive information. In addition, after the judgement it has become unclear whether search engine operators have a legal basis for indexing websites that contain special categories of data. We also discuss steps taken by Google to comply with the judgment. Section V contains concluding remarks.

## II. The Facts

In 1998, newspaper publisher La Vanguardia Ediciones SL published two announcements concerning a real-estate auction related to proceedings prompted by social security debts. In these announcements, Mario Costeja González was referred to as the owner of the real estate. The announcements were also published in the online version of the newspaper.[4] Google referred to La Vanguardia's pages if somebody searched for the name of Mario Costeja González.

González asked the publisher of the newspaper to delete the information, arguing that his financial troubles were long gone, and the information on the newspaper's website was of no relevance anymore. However, the publisher did not want to remove the information from the website, as the publication had been ordered by the Spanish Ministry of Labour and Social Affairs. González also contacted Google's office in Spain (Google Spain) to ask for the removal of links to the newspaper page in search results for his full name. Google Spain forwarded the request to Google Inc. in the United States (Google Inc.) as it was Google Spain's view that Google Inc. provided the search service.

In 2010, González filed a complaint with the Spanish data protection authority (DPA), arguing that La Vanguardia had to remove or rectify its publications. He also requested Google Inc. and Google Spain to be required to remove the references to his personal data on La Vanguardia's pages. The Spanish DPA rejected the complaint to the extent it related to La Vanguardia because the Ministry of Labour and Social Affairs had ordered the publication. However, the Spanish DPA upheld the complaint regarding Google Spain and Google Inc.

---

[3] See on the jurisdictional aspects of the judgment: Christopher Kuner, "The Right to be Forgotten and the Global Reach of EU Data Protection Law", 1 June 2014, available on the Internet at:
<www.concurringopinions.com/archives/2014/06/the-right-to-be-forgotten-and-the-global-reach-of-eu-data-protection-law.html> (last accessed on 22 August 2014).
[4] The original publication is available on the website of La Vanguardia at:
<http://hemeroteca.lavanguardia.com/preview/1998/01/19/pagina-23/33842001/pdf.html> (last accessed on 22 August 2014).

The Spanish DPA considered that search engine operators can be required to remove data from their indexes and to prevent future access to those data through their search engines.[5]

# III. The Judgment

*Material scope of the Data Protection Directive*

The first question that the CJEU deals with is whether a search engine's activity of indexing, automatically storing and referring to information available on the web must be seen as the "processing" of "personal data" in the sense of the Data Protection Directive.[6] Almost everything that can be done with personal data falls within the directive's definition of processing. It is thus not surprising that according to the CJEU, the activities described above entail the "processing" of personal data. Furthermore, the CJEU reaffirms that data protection law also applies to personal data that are already public. The CJEU adds that it is not relevant whether a search engine operator alters the data or not.[7]

Next, the CJEU answers the question of whether a search engine operator must be regarded as a "controller" regarding such processing activities. The directive defines a controller, in short, as the legal person or other body that alone or with others determines the purposes and means of the personal data processing.[8] The CJEU holds that a search engine operator determines the purposes and means of the processing at issue, and therefore qualifies as a controller.[9]

*Territorial scope of the Data Protection Directive*

The CJEU subsequently considers the territorial scope of the directive.[10] Article(4)(1)(a) of the directive reads: "[e]ach Member State shall apply the national provisions it adopts pursuant to this Directive to the processing of personal data where: the processing is carried out in the context of the activities of an establishment of the controller on the territory of the Member State; when the same controller is established on the territory of several Member States, he must take the necessary measures to ensure that each of these establishments complies with the obligations laid down by the national law applicable." Google claimed that no personal data processing relating to its search engine took place in Spain. Google argued that its processing was exclusively carried out by Google Inc. and that Google Spain's activities were limited to providing support to Google group's advertising activities, which were separate from the search engine service.[11]

The CJEU, however, says Google Spain is an establishment of Google Inc. in the sense of Article 4(1)(a), and that Google Inc.'s processing is "carried out in the context of the activities" of this establishment.[12] The CJEU holds that Article 4(1)(a) should be "interpreted as meaning that processing of personal data is carried out in the context of the activities of an establishment of the controller on the territory of a Member State, within the meaning of that provision, when the operator of a search engine sets up in a Member State a branch or subsidiary which is intended to promote and sell advertising space offered by that engine and which orientates its activity towards the inhabitants of that Member State."[13]

---

[5] See in more detail about the facts and national procedure: Google Spain, paras. 14-20; and Opinion of Advocate General Jääskinen in Case C-131/12 *Google Spain SL and Google Inc. v. Agencia Española de Protección de Datos and Mario Costeja González*, not yet published, at paras. 18-23.
[6] Google Spain, at paras. 20-21. See also Art. 2(a) and 2(b) of the Data Protection Directive.
[7] Google Spain, at para. 29-31.
[8] Art. 2(d) of the Data Protection Directive.
[9] Google Spain, at para. 33.
[10] Ibid., at para. 42.
[11] Ibid., at para. 51.
[12] Ibid. at para. 55.

***Right to have search results delisted***

In the EU Charter of Fundamental Rights, the right to privacy is protected by Article 7.[14] The Charter also recognises a separate right to the protection of personal data in Article 8. The requirements for fair and lawful data processing are laid down in more detail in the Data Protection Directive.[15] Article 6 of the directive says that personal data must be "adequate, relevant and not excessive" in relation to the processing purposes, and "kept in a form which permits identification of data subjects for no longer than is necessary" for those purposes.[16]

The directive grants data subjects several rights, two of which play a major role in the case: under certain circumstances, the data subject has the right to request erasure of personal data, and the right to object to processing. Under Article 12(b), Member States must guarantee every data subject the right to obtain from the controller, "as appropriate the rectification, erasure or blocking of data the processing of which does not comply with the provisions of this Directive, in particular because of the incomplete or inaccurate nature of the data."[17] Article 14(1)(a) grants the data subject the right, "at least in the cases referred to in Article 7 (e) and (f), to object on compelling legitimate grounds relating to his particular situation to the processing of data relating to him, save where otherwise provided by national legislation. Where there is a justified objection, the processing instigated by the controller may no longer involve those data."

The Spanish court asks advice about the scope of the data subject's rights. The question is whether the relevant provisions must "be interpreted as enabling the data subject to require the operator of a search engine to remove from the list of results displayed following a search made on the basis of his name links to web pages published lawfully by third parties and containing true information relating to him, on the ground that that information may be prejudicial to him or that he wishes it to be 'forgotten' after a certain time."[18]

The CJEU's notes that a search engine's activity may "affect significantly" the fundamental rights to privacy and to the protection of personal data.[19] After all, search results for a name provide "a structured overview of the information relating to that individual that can be found on the internet – information which potentially concerns a vast number of aspects of his private life and which, without the search engine, could not have been interconnected or could have been only with great difficulty – and thereby to establish a more or less detailed profile of him."[20]

As noted, Article 12(b) grants the data subject the right to block personal data processing if the processing does not comply with the directive. According to the CJEU, such "incompatibility [with the directive] may result not only from the fact that such data are inaccurate but, in particular, also from the fact that they are inadequate, irrelevant or excessive in relation to the purposes of the processing, that they are not kept up to date, or that they are kept for longer than is necessary unless they are required to be kept for historical, statistical or scientific purposes."[21] Furthermore, "even initially lawful processing of accurate data may, in the course of time, become incompatible with the directive where those data are no longer necessary in the light of the purposes for which they were collected or processed."[22] The CJEU adds: "[t]hat is so in particular where [personal data] appear to be inadequate, irrelevant or no longer relevant, or excessive in relation to those purposes and in the light of the time that has elapsed."[23]

---

[13] Ibid., at para. 60 and dictum. See generally on the territorial scope of data protection law: Kuner C, *Transborder Data Flows and Data Privacy Law (PhD thesis University of Tilburg, academic version)* (Kuner 2012), chapter VI and VII.
[14] Art. 7 of the Charter is phrased similarly as Art. 8 of the European Convention on Human rights.
[15] Google Spain, at para. 69.
[16] Art. 6(1)(c) and 6(1)(e) of the Data Protection Directive.
[17] Ibid., at para. 70.
[18] Ibid, at para. 89.
[19] Ibid, at para. 38.
[20] Ibid, at para. 80.

Therefore, says the CJEU, Article 12(b) and Article 14(1)(a) must "be interpreted as meaning that, when appraising the conditions for the application of those provisions, it should inter alia be examined whether the data subject has a right that the information in question relating to him personally should, at this point in time, no longer be linked to his name by a list of results displayed following a search made on the basis of his name, without it being necessary in order to find such a right that the inclusion of the information in question in that list causes prejudice to the data subject."[24]

According to the CJEU, "a fair balance should be sought" between the legitimate interests of searchers and the data subject's privacy and data protection rights.[25] The CJEU adds that the data subject's privacy and data protection rights "override, as a rule, not only the economic interest of the operator of the search engine but also the interest of the general public in having access to that information upon a search relating to the data subject's name."[26]

However, the CJEU immediately mentions an important caveat. The data subject's rights should not prevail "if it appeared, for particular reasons, such as the role played by the data subject in public life, that the interference with his fundamental rights is justified by the preponderant interest of the general public in having, on account of its inclusion in the list of results, access to the information in question."[27]

If, after a data subject request, the search engine operator decides not to delist search results based on name searches, the data subject may bring the matter before the data protection authority or the judicial authority so that it carries out the necessary checks and orders the controller to take specific measures accordingly.[28]

## IV.  Comments: Lack of Attention to Freedom of Expression

In cases where somebody wants search results to be delisted, a balance must be struck between the right to freedom of expression, the right to privacy, and the right to data protection.[29] The right to privacy and the right to freedom of expression are equally important. As the European Court of Human Rights puts it, "as a matter of principle these rights deserve equal respect."[30] Neither right is absolute, and balancing the rights should happen on a case-by-case basis, taking all circumstances into account.

---

[21] Ibid, at para. 92.
[22] Ibid, at para. 92.
[23] Ibid, at para. 93.
[24] Ibid, at para. 99 and dictum.
[25] Ibid, at para. 81.
[26] Ibid, at para. 99 and dictum.
[27] Ibid, at para. 99 and dictum.
[28] Ibid, at para. 77.
[29] A discussion of the search engine operator's right to freedom to conduct a business (Art. 16 of the Charter) falls outside the scope of this note. This right could also be considered in the "fair balance".

In contrast, the CJEU seems to view questions concerning the removal of search results primarily through a data protection lens. Furthermore, the CJEU does not refer to the detailed and nuanced case law of the European Court of Human Rights on balancing privacy and freedom of expression.[31] The approach of the CJEU is surprising because the removal of search results for name searches interferes with the right to freedom of expression of multiple parties.

First, publishers of information on the web have a right to freedom of expression. This right also protects the means of communication. As the European Court of Human Rights notes, "Article 10 [of the European Convention on Human Rights] applies not only to the content of information but also to the means of transmission or reception since any restriction imposed on the means necessarily interferes with the right to receive and impart information."[32] In the context of search engines, Van Hoboken notes that "[t]he protected interests of information providers under the right to freedom of expression can be best understood as the freedom to be included in the search engine's index and to find their way to an audience."[33]

The CJEU holds that a search engine operator may have to comply with a request for delisting from the search results, even if the original publication is lawful. True, the CJEU's judgment is limited to the removal of search results for a name search. For instance, an article "Public auction of 10 Summer Street", may have to be delisted for a search for "John Doe" who is mentioned in the article, but can still show up in the results when somebody searches for "10 Summer Street". Nevertheless, such a delisting limits the publisher's freedom of expression, because it makes the original publication harder to find – at least on the basis of a name search.[34] The Advocate General warned: "[t]his would entail an interference with the freedom of expression of the publisher of the web page, who would not enjoy adequate legal protection in such a situation, any unregulated 'notice and take down procedure' being a private matter between the data subject and the search engine service provider. It would amount to the censuring of his published content by a private party"[35]

Second, as Van Hoboken notes, "freedom of expression doctrine, as in the case of press freedom, should focus on protecting the way search engines contribute to the ideals underlying freedom of expression and the functioning of the networked information environment as a whole."[36] He argues that, in principle, search engines have a freedom of expression claim: "under Article 10 ECHR [European Convention on Human Rights], the search engine should be able to claim protection for its publication of references on its website as well as the process of crawling that makes it possible to offer a search engine in the first place."[37] In contrast, the CJEU is silent on the right to freedom of expression of search engine operators and of the original publishers of information.

---

[30] ECtHR, Axel Springer AG v. Germany, app. no. 39954/08 (7 February 2012), at para. 87. See similarly: ECtHR, Von Hannover v. Germany, app. nrs. 40660/08 and 60641/08 (7 February 2012), at para. 100; ECtHR, Węgrzynowski and Smolczewski v. Poland, app. no. 33846/07 (16 July 2013), at para. 56. The European Convention on Human Rights does not contain a right to data protection, and the European Court of Human Rights largely includes data protection rights in the right to respect for private life (Art. 8 of the Convention).
[31] More generally, scholars have criticised the CJEU's lack of attention to the case law of the European Court of Human Rights. See: Grainne De Burca, "After the EU Charter of Fundamental Rights: The Court of Justice as a Human Rights Adjudicator?", 10 *Maastricht Journal of European and Comparative Law* (2013), pp. 168 *et sqq.*
[32] ECtHR, Autronic AG v. Switzerland, app. no. 12726/87 (22 May 1990), at para. 47.
[33] Joris van Hoboken, *Search Engine Freedom. On the Implications of the Right to Freedom of Expression for the Legal Governance of Web Search Engines* (Alphen aan den Rijn: Kluwer Law International, 2012), at. p. 350.
[34] After the judgment, Google not only removes results for name searches, but also for queries consisting of a combination of a name and other search terms. See: Letter from Google to Article 29 Working Party, 31 July 2014, available on the Internet at: <http://goo.gl/vQRE3B> (last accessed on 22 August 2014), answer to question 14.
[35] Opinion of Advocate General Jääskinen in Google Spain, at para. 134 (internal footnotes omitted).
[36] Van Hoboken, *Search Engine Freedom*, supra note 35, at p. 351.

Third, the right to freedom of expression, as protected in the Charter and the European Convention on Human Rights, includes the right "to receive and impart information and ideas without interference by public authority and regardless of frontiers."[38] Delisting search results may thus also interfere with the public's right to search for and find information on the web. As the European Court of Human Rights puts it, "the public has a right to receive information of general interest,"[39] and "the internet plays an important role in enhancing the public's access to news and facilitating the sharing and dissemination of information generally."[40]

While the removal of search results interferes with the right to freedom of expression of multiple parties, the judgment contains no explicit reference to this fundamental right. Staying close to the text of the directive, the CJEU refers to the "interests" of searchers in having access information.[41] This remark could be seen as an implicit reference to the right to receive information. However, as noted, the right to freedom of expression protects not only the searcher's interest in having access to information, but also the original publisher and, arguably, the search engine operator.

The CJEU does say, correctly, that "a fair balance" must be struck between the legitimate interests of searchers and the data subject's right to privacy and data protection.[42] But the CJEU adds that the data subject's privacy and data protection rights "override, as a rule, not only the economic interest of the operator of the search engine but also the interest of the general public in having access to that information upon a search relating to the data subject's name."[43] With that "rule", the CJEU takes a different approach than the European Court of Human Rights, which says that freedom of expression and privacy have equal weight.

The CJEU does add that in certain circumstances the interests of searchers should outweigh the data subject's rights, for instance because of "the role played by the data subject in public life."[44] This remark brings to mind the approach of the European Court of Human Rights when balancing privacy and the freedom of expression. That Court takes into account, among other factors, how well-known the person is about whom a publication speaks, and whether a publication contributes to a debate of general interest.[45] The CJEU's remark about the data subject's role in public life could mitigate the chance that a search engine operator delists a search result that contributes to the public debate. Nevertheless, the CJEU's "rule" that privacy and data protection rights override the right to receive information of searchers (the "interests" of searchers according to the CJEU) implies an unfortunate departure from the case law on balancing by the European Court of Human Rights.

---

[37] Ibid., at p. 228. Also see this U.S. court decision: *Search King, Inc. v. Google Technology, Inc.*, 2003 WL 21464568 (W.D. Okla. 2003).
[38] Art. 11(1) of the Charter of Fundamental Rights of the European Union; Art. 10 of the European Convention on Human Rights.
[39] ECtHR, Társaság a Szabadságjogokért v. Hungary, app. no. 37374/05 (14 April 2009), at para. 26.
[40] ECtHR, Fredrik Neij and Peter Sunde Kolmisoppi v. Sweden, app. no. 40397/12 (19 February 2013), at p. 9.
[41] The Directive mentions "interests" inter alia in article 7(f), the balancing provision. Below, under the heading "Special categories of personal data", we discuss that provision.
[42] Google Spain, at para. 81
[43] Ibid., dictum under nr. 4. See also para. 97 and 99.
[44] Ibid., dictum under nr. 4.
[45] See e.g. ECtHR, Von Hannover v. Germany, app. nrs. 40660/08 and 60641/08 (7 February 2012), at para. 108-113; ECtHR, Axel Springer AG v. Germany, app. no. 39954/08 (7 February 2012), at para. 89-95.

*Problems with private ordering*

Search engine operators do not necessarily present a neutral picture of the information available on the web, as they present a (algorithmically) curated list of search results.[46] It could thus be argued that Google is partly responsible for creating a problem for González. Without Google Search, people, including possibly prospective clients of González, would probably not have been confronted with an out-dated publication from about him. Arguably the image of González shown in the search results was partly created by Google. Perhaps the CJEU thought along the following lines: Google has created a problem for González and it should bear the cost of solving that problem. As the UK Information Commissioner reportedly told the BBC: "the polluter pays, the polluter should clear up."[47]

However, seeing search engine operators as controllers regarding the processing of personal data on third-party source web pages also implies assigning these operators the often complicated task of balancing the fundamental rights at stake. There may be clear-cut cases where privacy should prevail over freedom of expression, and in which it is sensible that a search engine operator delists a search result. However, in more difficult cases, search engine operators may not be the most appropriate party to balance the fundamental rights involved.

The Advocate General did not deem a search engine operator to be a controller regarding the processing of personal data on third-party source web pages, as long as the operator's activities are limited to "merely supplying an information location tool."[48] He discouraged the CJEU "from concluding that these conflicting interests could satisfactorily be balanced in individual cases on a case-by-case basis, with the judgment to be left to the internet search engine service provider."[49] Moreover, he warned that such a conclusion could lead to unmanageable number of delisting requests to search engine operators, and that operators might automatically withdraw search results.[50] Already in 2008, the Article 29 Working Party, an advisory body in which national Data Protection Authorities cooperate, recognised the important role of search engines for freedom of expression. The Working Party did not see search engine operators as primary controllers when they act "purely as an intermediary".[51]

The private ordering system created through the Google Spain judgment resembles the system we know from the E-Commerce Directive's liability exemption for hosting services, where Internet intermediaries, under threat of incurring liability, are pressured to decide on the lawfulness of the third-party content they help to communicate. To benefit from the liability exemption, those intermediaries need to remove content "expeditiously" once they acquire knowledge of the unlawfulness of the content.[52] This regime creates an incentive for intermediaries to systematically take down content after complaints, which may interfere with the freedom of expression of those communicating on the internet.[53]

---

[46] On this issue see e.g.: James Grimmelmann, "Speech Engines", 98 Minnesota Law Review (2014), pp. 868 *et sqq*; Van Hoboken, *Search Engine Freedom*, *supra* note 35, at p. 43-50; p. 189-215.
[47] Kelly Fiveash, "ICO: It's up to Google the 'polluter' to tidy up 'right to be forgotten' search links", 24 July 2014, available on the Internet at:
<www.theregister.co.uk/2014/07/24/ico_chief_says_google_needs_to_tidy_up_right_to_be_forgotten_requests_as_search_engines_meet_brussels_officials>. An economist might use the phrase "negative externalities" to characterize the costs that Google imposed on Gonzalez (in the form of reputational harm), and on DPAs and courts (in the form of more work).
[48] Opinion of Advocate General Jääskinen in Google Spain, at para. 84. He added that a search engine operator would be a controller if it "index[ed] or archive[d] personal data against the instructions or requests of the publisher of the web page (ibid., at para. 138).
[49] Ibid., at para. 133.
[50] Ibid., at para. 133.
[51] Article 29 Working Party, "Opinion 1/2008 on data protection issues related to search engines", 4 April 2008, at p. 14.

A general problem of private ordering by online service providers through notice and takedown mechanisms is the lack of transparency of their decisions. Without clarity on which results have been delisted, members of the public have limited ability to know the extent to which their freedom to receive information has been interfered with. Google is one of few Internet companies that reports on removal requests from copyright owners and removal of search results. These reports show interesting information, such as who are the "top copyright owners" requesting removals. But as these reports provide aggregated data, it is impossible to see what exact content has been removed.[54] The U.S. Chilling Effects website does provide detailed removal requests, but the transparency provided on this website is dependent on people sharing removal requests that they received.[55] Below we will see that a search engine operator would face additional difficulties if it wanted to report on removals after data subject requests.

*Media exemption*

The CJEU could have given search engine operators the possibility to benefit from the directive's media exemption. Under Article 9 of the directive, Member States can provide exemptions from parts of the directive (including Article 12 and 14), if the processing is "carried out solely for journalistic purposes" as long as those exemptions "are necessary to reconcile the right to privacy with the rules governing freedom of expression."

In the *Satamedia* case, the CJEU has interpreted this media exemption very broadly. The CJEU said that in principle, a provider of a text-messaging service that enables people to obtain information about any individual's income and assets could rely on the media exemption.[56] In this light, it is not too far-fetched to allow search engine operators to benefit from the media exemption. Indeed, in 2008, the Article 29 Working Party has suggested that, in order to strike the right balance between freedom of expression and data protection law, search engines should be able to benefit from the media exemption.[57] But in the *Google Spain* case, the CJEU says a search engine operator cannot benefit from the media exemption.[58]

---

[52] Art. 14(1)(b) of the E-Commerce Directive (European Parliament and Council Directive 2000/31/EC on certain legal aspects of information society services, in particular electronic commerce, in the Internal Market, OJ 2000 L 178/1).
[53] Rosa Julià-Barceló and Kamiel J Koelman, "Intermediary Liability In The E-Commerce Directive: So Far So Good, But It's Not Enough" 16 Computer Law & Security Report (2000), pp. 231 *et sqq.* For similar problems in the United States: Jennifer Urban and Laura Quilter, "Efficient process or 'Chilling Effects'? Takedown Notices under Section 512 of the Digital Millenium Copyright Act", 22 Santa Clara Computer & High Tech. LJ (2005), pp. 621 *et sqq.*, at p. 638; Wendy Seltzer, "Free Speech Unmoored in Copyright's Safe Harbor: Chilling Effects on the DMCA on the First Amendment" 24 Harvard Journal of Law & Technology (2010), pp. 171 *et sqq.*, at pp. 224–225.
[54] Google, "Transparency Report", available on the Internet at: <www.google.com/transparencyreport> (last accessed on 22 August 2014).
[55] Chilling Effects Clearinghouse, available on the Internet at: <www.chillingeffects.org> (last accessed on 22 August 2014).
[56] Case C-73/07, Tietosuojavaltuutettu v. Satakunnan Markkinapörssi Oy and Satamedia Oy, [2008] ECR I-09831, dictum.
[57] Article 29 Working Party, "Opinion 1/2008 on data protection issues related to search engines", 4 April 2008, at p. 13.
[58] Google Spain, at para. 85. In the original language of the case (Spanish), the CJEU says Google cannot benefit from the media exception: "ése no es el caso en el supuesto del tratamiento que lleva a cabo el gestor de un motor de búsqueda" ("this is not the case in the event of the processing [of personal data] that the operator of a

*Special categories of personal data*

The CJEU's choice to see a search engine operator as the data controller regarding the processing of personal data on third party web pages has another remarkable implication. The directive only allows personal data processing if the controller can rely on one of the six legal bases for processing listed in Article 7.[59] The CJEU suggests that, for the processing at issue, a search engine operator can rely on the balancing provision (Article 7(f)).[60] In short, a controller can rely on the balancing provision when its legitimate interests, or those of a third party, outweigh the fundamental rights of the data subject.

However, the processing of "special categories of data", such as personal data regarding racial or ethnic origin, political opinions, religious beliefs, and data concerning health or sex life, cannot be based on the balancing provision. Unless a specified exception applies, the processing of special categories of data is prohibited, or, depending on the national implementation law, only allowed after the data subject's explicit consent.[61]

A search engine operator's processing may involve indexing of many web pages that include special categories of data. For instance, a web page might include pictures of people that show their skin colour and racial origin. Similarly, a member list on the website of a Catholic choir arguably contains data indicating a person's religion.

A search engine operator would need the data subject's explicit consent for indexing pages that include special categories of data. However, this requirement would render the operation of a search engine practically impossible. True, there are exceptions to the in-principle prohibition of processing special categories of data, such as exceptions for churches and for the medical sector.[62] Furthermore, the in-principle prohibition does not apply when "the processing relates to data which are manifestly made public by the data subject."[63] In some cases this exception might apply, for instance if the members of the Catholic choir gave their explicit consent to the choir to publish their names online. But in many cases, such as with pictures of people on the web, the relevant data subjects may not have made those data public themselves. Hence, it seems that in many cases search engine operators cannot rely on any of the exceptions.

In sum, it seems that seeing a search engine operator as a controller regarding the processing of personal data on third party web pages implies that the operator's practices are partly illegal. To avoid such a consequence, and taking into account the important function of search engines, the Advocate General did not want to qualify the search engine operator as a controller regarding the processing of personal data on third party web pages.[64] However, the CJEU ignores the problem of a lack of legal basis for the indexing of web pages that contain special categories of data.

---

search engine carries out"). As the original language of the case, Spanish is the authentic language of the judgment (see Article 41 of the Rules of Procedure of the Court of Justice of the European Union). The Dutch and German versions of the judgment are in line with the Spanish text. The English (and the French) version say Google "does not appear" to be able to benefit from the media exception, and thus incorrectly imply that Google might benefit from the exception.
[59] Art. 8(2) of the Charter of Fundamental Rights of the European Union, Art. 7 of the Data Protection Directive.
[60] Google Spain, at para. 73.
[61] Art. 8 of the Data Protection Directive.
[62] Art. 8(2)(d) and Art. (8)(3) of the Data Protection Directive.
[63] Art. 8(2)(e) of the Data Protection Directive.
[64] Opinion of Advocate General Jääskinen in Google Spain, at para. 90.

*Google's implementation of the judgment*

The Article 29 Working Party has met with search engine operators to discuss the implementation of the judgment. In answers to the Working Party, Google explained it received 91.000 requests involving more than 300.000 URLs as of 18 July 2014. Google delisted about half of the URLs upon request. For 32% it decided not to delist the URL. For 15% Google asked for more information from the requester.[65] As an aside, Google has some experience with receiving requests to delist search results. Google receives around 6 to 8 million removal requests per week for pages allegedly containing copyright infringing material, and it removes most of those pages.[66]

When a search engine operator wants to report on the delisting of search results based on data protection law, it faces difficulties. Specific transparency reports on such removals, including the reasons why the results have been removed, could interfere with the requester's privacy rights.[67] The UK Information Commissioner notes that in some cases it might not be wise to inform the original publisher about a removal request, for instance in the case of "hate sites of various sorts". In such cases, "informing the content publisher could exacerbate an already difficult situation and could in itself have a very detrimental effect on the complainant's privacy."[68]

If Google delists a search result, it informs the relevant website publisher. Google tells the website publisher which URL it delisted, but does not disclose who submitted the request or other details about the request.[69] This practice informs the publisher about the removal, and gives that publisher a clue to find out why their page has been removed from certain search results. For instance, a website publisher that is notified about the delisting of a page concerning politician X, can check whether the politician caused the removal by searching for the politician's name, and seeing if the page shows up. If the publication does not show up, and the publisher does not agree with the delisting, the publisher can ask Google to re-evaluate the delisting.[70] If a publisher is informed about a removal request such as the one regarding the 16-year old auction of the real estate of González, a publisher might decide that the delisting is reasonable.

At the time of writing, Google displays on its European pages the somewhat vague notice that "Some results may have been removed under data protection law in Europe" for search queries that, according to Google, are name searches. Google thus shows this notice on pages with search results for most names, and not just on those pages that have been affected by a delisting.[71] Google explains to the Article 29 Working Party: "[t]he notification is intended to alert users to the possibility that their results for this kind of query may have been affected by a removal, but not to publicly reveal which queries were actually affected."[72] But this practice gives the public limited insight into which search results are delisted, leaving the public with no oversight over how their right to access of information is limited.

---

[65] Letter from Google to Article 29 Working Party, *supra* note 36, answer to question 23.
[66] For instance, Google receives around 6 to 8 million removal requests per week for pages allegedly containing copyright infringing material. Google indicates that it removed 97% of search results specified in requests that it received between July and December 2011. See: Google, "Transparency Report", available on the Internet at: <www.google.com/transparencyreport/removals/copyright/> (last accessed on 22 August 2014); and Google, "FAQ", available on the Internet at:
<www.google.com/transparencyreport/removals/copyright/faq/#compliance_rate> (last accessed on 22 August 2014).
[67] Ibid., answer to question 20. See also David Drummond, "We need to talk about the right to be forgotten", 10 July 2014, available on the Internet at: <www.theguardian.com/commentisfree/2014/jul/10/right-to-be-forgotten-european-ruling-google-debate> (last accessed on 22 August 2014).
[68] David Smith, Deputy Commissioner and Director of Data Protection, "Update on our response to the European Google judgment", 7 August 2014, available on the Internet at:
http://iconewsblog.wordpress.com/2014/08/07/update-on-our-response-to-the-european-google-judgment/
[69] Letter from Google to Article 29 Working Party, *supra* note 36, answer to question 6.
[70] A difficulty here is that such a name search query may give hundreds, if not thousands, of search results.
[71] Google, "FAQ", available on the Internet at: <www.google.co.uk/intl/en/policies/faq> (last accessed on 22 August 2014).

Many questions regarding the judgment are still open. For instance, the judgment applies to "the operator of a search engine". Which types of companies are within the scope of the judgment? Twitter has a search function for tweets; does that make Twitter a search engine operator? Are newspaper websites with a search function within the scope of the judgment? And how to deal with rapidly changing webpages, such as a Wikipedia page? The removal of such a page for a name search may quickly no longer be justified. And if data subjects turn to data protection authorities if a search engine operator did not comply with their requests, will these authorities be able to cope with the workload? Is it reasonable that DPAs and courts – and indirectly the general public – incur costs for a problem that is arguably created, in part, by search engine operators? What will be the judgment's effect on discussions in Brussels about the right to erasure in the proposal for a Data Protection Regulation (misleadingly called a "right to be forgotten" in earlier drafts)?[73]

## V. Concluding Remarks

Contrary to what some press reports suggest, the Google Spain judgment does not create a right to be forgotten. Rather, the CJEU holds that people, under certain conditions, have the right to have search results for their name delisted. This right can also extend to lawfully published information.

We have argued that the CJEU did not give enough attention to the right to freedom of expression. The CJEU suggests that privacy and data protection rights override, "as a rule", the public's right to receive information (referred to as an "interest" rather than a right by the CJEU). That "rule" is an unfortunate departure from the doctrine developed by European Court of Human Rights. The CJEU should have given equal weight to the right to freedom of expression (which includes the right to receive information), the right to privacy, and the right to data protection. Furthermore, by seeing search engine operators as controllers regarding the processing of personal data on third party web pages, the CJEU assigns these operators the potentially complicated task of balancing the fundamental rights at stake. But especially in difficult cases, a search engine operator may not be the most appropriate party to conduct such balancing acts.

Additionally, after the decision, it seems that in some situations search engine operators do not have a legal basis for indexing websites that contain special categories of data. In principle, a search engine operator would need the data subject's explicit consent for indexing pages that include special categories of data. Such a requirement would render the activities of a search engine operator practically impossible.

Finally we add some thoughts about the broader subject of out-dated information on the web that interferes with people's privacy. Such situations present complicated questions, where various rights may be in conflict. Unfortunately, there are no easy answers. But policymakers can seek solutions without trying to solve all issues by putting obligations on search engine operators. For instance, the Spanish Ministry of Labour and Social Affairs should rethink the obligation of publishing public notices of real-estate auctions connected with proceedings for the recovery of social security debts. Perhaps such notices should only be indexed in search engines for a few years.[74] And countries where judicial rulings are published on the web without proper anonymisation should re-evaluate that practice, now that search engines make such information easily retrievable. In sum, different situations call for different solutions. In the most difficult cases, it would probably be best if the courts would balance privacy and freedom of expression. In such cases courts should give equal weight to these fundamental rights.

---

[72] Letter from Google to Article 29 Working Party, *supra* note 36, answer to question 20
[73] See Art. 17 of the proposed Data Protection Regulation (COM(2012) 11 final, 2012/0011 (COD), 25 January 2012). The term "right to be forgotten" implies that someone has the right to have others forget something. It is therefore a misleading term for a right to have one's personal data deleted under certain conditions.

* * *

---

[74] The original publisher could use a do-not-index code for search engines, for instance in a robots.txt-file. (See Opinion of Advocate General Jääskinen in Google Spain, at para. 41).